\documentclass[aps,prd,amsmath]{revtex4}
\usepackage{graphicx}

\begin{document}

\title{Comments on "Light deflection by Damour-Solodukhin wormholes and Gauss-Bonnet theorem, Phys. Rev. D 98, 044033 (2018) by Ali \"{O}vg\"{u}n"}
\author{Amrita Bhattacharya}
\email{amrita_852003@yahoo.co.in}
\affiliation{Department of Mathematics, Kidderpore College, 2, Pitamber Sircar Lane, Kolkata 700023, WB, India}
\author{Ramis Kh. Karimov}
\email{karimov_ramis_1992@mail.ru}
\affiliation{Zel'dovich International Center for Astrophysics, Bashkir State Pedagogical University, 3A, October Revolution Street, Ufa 450008, RB, Russia}
\date{\today}

%\begin{abstract}

%\end{abstract}

%\pacs{}
\maketitle

%%%%%%%%%%%%%%%%%%%%%%%%%%%%%%%%%%%%  DATE  %%%%%%%%%%%%%%%%%%%%%%%%%%%%%%%%%%%%

In the article cited above, \"{O}vg\"{u}n \cite{Ovgun:2018aa} calculated the leading order
deflection angle of light by the static Damour-Solodukhin wormhole \cite{Damour:2007ab} using
its redefined static and Kerr-like forms proposed by Bueno \textit{et al}
\cite{Bueno:2018bb}. Because of the importance of wormholes, especially with regard to
strong field lensing [4,5] and gravitational waves \cite{Damour:2007ab,Bueno:2018bb,Cardoso:2016cf,Cardoso:2017cp,Konoplya:2016kz,Nandi:2017is,Vokel:2018vk}, we decided to
take a closer look at the work in \cite{Ovgun:2018aa}.

The purpose of this Comment is to point out that there are conceptual and
mathematical flaws in the analysis in \cite{Ovgun:2018aa}. First, the coefficients  $%
\overline{a}$ and $\overline{b}$ derived by the author are incorrect as $%
\overline{a}$ in \textit{Eqn.(2.31) }has a wrong sign\textit{\ }(probably a
typo) but $\overline{b}$ in \textit{Eqn.(2.32)} is in error. Second, the
deflection angle in \textit{Eqn.(3.10)} in the Kerr-like case is plainly
incorrect.

\"{O}vg\"{u}n takes the redefined form of the metric given in \cite{Bueno:2018bb}, viz.,

\begin{equation}
ds^{2} = -\left(1-\frac{2M}{r}\right)c^{2}dt^{2} + \left[1-\frac{2M(1+\lambda^{2})}{r}\right]^{-1}dr^{2} + r^{2}d\theta^{2}+r^{2}\sin ^{2}\theta d\varphi^{2},
\end{equation}%
where $\lambda $ is a constant parameter. The strong field deflection angle
in the static case due to Bozza is \cite{Bozza:2002bz}

\begin{equation}
\widehat{\alpha }(\theta )=-\overline{a}\log \left( \frac{u}{u_{m}}-1\right)+\overline{b}.
\end{equation}

The author obtains the Bozza coefficients $\overline{a}$ and $\overline{b}$
of the strong field deflection angle for the static case as follows (Eqs.(2.31)-(2.32))%
\begin{eqnarray}
\overline{a} &=&-\frac{1}{\sqrt{1-2\lambda^{2}}}, \nonumber \\
\overline{b} &=&\frac{\log \left( 6\right) \sqrt{-2\lambda ^{2}+1}}{\log
(10)2\lambda ^{2}-1}+\left( r_{m}-\pi \right).  \nonumber
\end{eqnarray}%
The correct expressions of the coefficients for the metric (1) however work
out to%
\begin{eqnarray}
\overline{a} &=&\frac{1}{\sqrt{1-2\lambda ^{2}}}, \\
\overline{b} &=&\frac{\log \left( 6\right) }{\sqrt{1-2\lambda ^{2}}}+\left(b_{R}-\pi\right),  \\
b_{R} &=&\int_{0}^{1}g(z,r_{m})dz, \\
g(z,r_{m})&=&2\left[ \frac{\sqrt{3+6z}}{z\sqrt{(3-2z)\{1+2z-2(1-z)\lambda
^{2}\}}}-\frac{1}{z\sqrt{1-2\lambda ^{2}}}\right] .
\end{eqnarray}%
Observe that the first part in $\overline{b}$ in \textit{Eqn.(2.32)} is
different from the corresponding part in the expression in (4). The second
part $\left( r_{m}-\pi \right) $ in \textit{Eqn.(2.32) }depending on $M$ via
$r_{m}=3M$ becomes dimensionful differing from $\left( b_{R}-\pi \right) $
in the expression in (4). However, note that the Bozza coefficients are
dimensionless, which is achieved in (6) since at $r_{m}=3M$ the expression
for $g(z,r_{m})$ becomes independent of $M$, and consequently $\left(
b_{R}-\pi \right) $ becomes dimensionless, as required. The illustrative
table below gives a glimpse of the variation of coefficients with $\lambda $
allowing a comparison with the known Schwarzschild values:

\begin{table}[!ht]
\begin{tabular}{|c|c|c|}
  \hline
  $\lambda$  & $\overline{a}$ & $\overline{b}$ \\
  \hline
  $0$ (Sch)  & $1.0000$       & $-0.4002$ \\
  $0.1$      & $1.0101$       & $-0.4131$ \\
  $0.2$      & $1.0425$       & $-0.4573$ \\
  $0.3$      & $1.1043$       & $-0.5536$\\
  \hline
\end{tabular}
\caption{Strong field lensing coefficients for Damour-Solodukhin wormhole.}
\label{Tab:Efficiency}
\end{table}

Next, \"{O}vg\"{u}n [1] obtained the leading order deflection in the
Kerr-like case as (Eqn.(3.10))%
%\begin{equation}
$$\widehat{\alpha }=\frac{2M(\lambda ^{2}+2)}{b}\pm \frac{4Ma}{b^{2}},$$
%\end{equation*}%
meaning that the term in $1/b^{2}$ is symmetric on both sides of the lens.
This result is incorrect. Conceptually, the deflection produced by the
presence of spin $a$ (inducing frame dragging) explicitly depends on whether
the light motion takes place in the direction of spin or opposite to it \cite{Iyer:2009ih}
(see also \cite{Iyer:2010yy}). Compared to the static case, the deflection is greater for
direct and smaller for retrograde orbits resulting in the \textit{loss} of
symmetry about the spin axis. The higher the spin, the tighter is the
winding of the direct orbit and more unwinding of the retrograde orbits.
Mathematically, the exact bending angle for the Kerr case ($\lambda =0$) has
been worked out by Iyer and Hansen \cite{Iyer:2009ih} and in the weak field it expands as
\begin{equation}
\widehat{\alpha }=\frac{4M}{b}+\left[ \frac{15\pi }{4}-4s\left( \frac{a}{M}%
\right) \right] \left( \frac{M}{b}\right) ^{2}+\left[ \frac{128}{3}-10\pi
s\left( \frac{a}{M}\right) +4\left( \frac{a}{M}\right) ^{2}\right] \left(
\frac{M}{b}\right) ^{3}+...
\end{equation}%
where $s=+1$ for direct orbits and $-1$ for retrograde orbits. The loss of
symmetry in the coefficient of $\left( \frac{M}{b}\right) ^{2}$ (as well as
in higher order terms) is quite evident. The limiting Kerr value $\lambda =0$
in \textit{Eqn.(3.10)} does not reproduce the second term in Eqn.(7).

%\section*{Acknowledgments}

%The reported study was funded by RFBR according to the research project No. 18-32-00377.

\end{document}